# Epitaxial Displacement-etched GaAsSb Quantum Dots Emitting at Telecom Wavelengths

Sadhvikas Addamane[1*], Sanghyeok Park[1], Quinn T. Campbell,[2] Evan M. Anderson,[2] Thomas Rotter[3], Darryl Shima[3], Julia Deitz[2], Paul Kotula[2], Ganesh Balakrishnan[3], Oleg Mitrofanov[4], Igal Brener[1] and Prasad P. Iyer[1*]

*[1]Center for Integrated Nanotechnologies, Sandia National Laboratories, Albuquerque, NM 8712, USA*
*[2]Sandia National Laboratories, Albuquerque, NM 8712, USA. [3]Center for High Technology Materials, University of New Mexico, Albuquerque, NM, 87106. [4]University College London, Electronic and Electrical Engineering, London WC1E 7JE, U.K.*
*saddama@sandia.gov, *ppadma@sandia.gov

**Abstract:** Semiconductor quantum dots (QDs) are essential components for integrated photonic and quantum information systems, yet current epitaxial growth techniques face significant limitations in complexity of growth, material selection, and density tunability. Here we report a displacement etch technique that enables controlled formation of GaAsSb QDs with emission at telecommunications wavelengths. Our approach uses arsenic molecular flux ($As_2$) to react with antimonide surfaces through thermodynamically favorable displacement reactions, creating three-dimensional quantum confined nanostructures without requiring strain-driven growth or complex droplet-based processes. Density functional theory shows the two competing As-for-Sb exchange pathways to be nearly equivalent in both reaction energy and activation barrier, which places morphological control in the subsequent adatom kinetics rather than in the surface reaction itself. A single process parameter, the $As_2$ exposure, sets the QD areal density across three orders of magnitude, from ~$10^2$ to ~$10^5$ $\mu m^{-2}$, covering both isolated emitters and dense gain media. Structural analysis reveals well-defined GaAsSb QDs with a mixed GaAs/GaSb composition and no extended defects. Temperature-dependent photoluminescence measurements confirm emission at ~1550 nm (~0.8 eV) within the telecommunications C-band, with characteristic QD-like temperature dependence. Because the QDs form on a lattice-matched GaSb surface in a single growth step, the technique offers a practical route to C-band emitters built into the GaSb/AlGaSb device family already used for mid-infrared sources and detectors.

## Introduction

Precise control of light emission properties through epitaxial growth of quantum confined nanostructures has emerged as a cornerstone of both integrated photonic sources and quantum information sciences.[1–3] Semiconductor quantum dots (QDs) have generated light with tailored spectral, temporal, and quantum properties, from single-photon sources enabling secure quantum communication to high-performance lasers operating at telecommunications wavelengths.[4] Quantum confinement effects in these zero-dimensional nanostructures enable precise engineering of emission wavelengths, linewidths, and carrier dynamics, making them indispensable components in next-generation photonic and quantum technologies.[5] The epitaxial growth of light-emitting semiconductor QDs with spatial density control has enabled the development of applications ranging from high-brightness single-photon sources[6–8] at low densities to telecommunications lasers at high densities.[4]

Currently, two major epitaxial growth techniques dominate the field of III-V semiconductor QD fabrication, each successful but each with limitations that prevent lattice-matched material selection and wide-range density control.[9,10] The Stranski–Krastanov (S-K) growth mode exploits lattice mismatch-induced strain to drive three-dimensional island formation, exemplified by the widely studied InAs QDs on GaAs substrates.[11] This approach has enabled decades of advances in QD lasers, single-photon emitters, and photodetectors, benefiting from the maturity of the epitaxial growth process. However, S-K growth is fundamentally constrained by the requirement for specific lattice-mismatched material combinations, restricting accessible emission wavelengths and preventing the use of potentially superior material systems for telecommunications applications.[12] The constraint is particularly challenging in the telecommunications C-band, where InAs QDs on GaAs emit near 1.3 μm[13] and reach 1.55 μm only through metamorphic buffers or strain-reducing layers,[14,15] leaving C-band QD emitters largely confined to the less-mature InP platform.[3,16] The self-assembled nature of S-K growth also offers limited independent control over QD size and areal density, introducing inherent randomness that results in inhomogeneous broadening and compromises device uniformity.[13,17,18] The second established technique, droplet epitaxy and its derivative local droplet etching (LDE), attempts to overcome material limitations by using metal droplets to create droplet-etched holes that are subsequently filled with desired semiconductor materials.[10] While LDE offers greater flexibility in material selection,[19–22] it introduces significant challenges including complex multi-step growth recipes,[10] difficulties in complete droplet material removal,[23] compositional variation,[24] and generally lower spatial density compared to S-K QDs.[25,26] Most critically, neither technique provides the combination of material flexibility, processing simplicity, and wide-range density control needed for emerging quantum photonic applications.

Here, we report an epitaxial approach that uses displacement etch reactions to form GaAsSb QDs and addresses the above-mentioned limitations through simple processing and wide-range density control. Our technique employs arsenic molecular flux ($As_2$) to react with antimonide surfaces,[27–29] creating a controlled displacement reaction that forms three-dimensional quantum confined structures without requiring lattice-mismatched growth or complex droplet-based processes. The technique yields QD areal densities controllable over three orders of magnitude from $\sim 10^2$ to $\sim 10^5$ $\mu m^{-2}$, spanning the range needed for both collective emission devices and isolated single-photon sources. The practical advance is that one in-situ monitorable process parameter, the $As_2$ exposure, sets the areal density across the range that device applications demand, in a single growth step, on a lattice-matched substrate, and without the multi-step droplet recipes or the strain-driven material restrictions that constrain the established routes. In addition, the platform used here is technologically relevant. GaSb and AlGaSb are the host materials for mid-infrared sources and detectors,[30,31] and lattice-matched C-band emitters on that platform are not routinely available, so QDs that grow directly into a GaSb/AlGaSb stack removes an integration step. Through electron-microscopy based structural analysis and temperature-dependent photoluminescence measurements, we demonstrate the areal density control and telecommunications wavelength emission from GaAsSb QDs at ~1550 nm, positioning these QDs as promising candidates for fiber-optic quantum communication systems and integrated photonic devices.

## Experimental Section

**Epitaxial growth:** The epitaxial growth described in this article was carried out using molecular beam epitaxy (MBE) on (100) GaSb substrates. The substrate temperature was measured using a pyrometer and the growth rates were calibrated and monitored using reflection high-energy electron diffraction (RHEED). The MBE systems were equipped with valved cracker sources for both arsenic and antimony, affording precise control over both fluxes. The surface oxide on GaSb substrates was first thermally desorbed at ~540 °C for 15 min. The substrate temperature was then adjusted to 510–520 °C and a 100 nm GaSb smoothing layer was grown under standard GaSb conditions, at a Ga growth rate of 0.35 ML/s and an Sb:Ga beam equivalent pressure (BEP) ratio of ~5, to provide an atomically flat starting surface. Following this, the substrate temperature was adjusted for the $As_2$ etch process (if needed) and the GaSb surface was exposed to a controlled $As_2$ flux. An ion-gauge beam flux monitor was used to measure the BEPs; the $As_2$ BEPs were in the range of $7\times10^{-7}$ and $1\times10^{-5}$ Torr and the exposure time was varied from 30 s to over 600 s. The $As_2$ etch was followed by a 300 s Sb soak to aid the formation of the nanostructures. The structure is completed with the growth of a GaSb cap on top. For atomic force microscopy (AFM) samples, the etch process ($As_2$ etch followed by 300 s Sb soak) was repeated at the terminating top GaSb surface. The photoluminescence (PL) samples differ from the AFM samples only in the surrounding barrier: the QDs are formed by etching a 4 nm GaSb quantum well grown at the center of a 200 nm $Al_{0.5}Ga_{0.5}Sb$ layer, so the displacement etch always acts on a GaSb surface. The $Al_{0.5}Ga_{0.5}Sb$ serves only to confine carriers, and the QDs are therefore GaAsSb.

**Structural characterization:** Nanostructure dimensions were measured mainly using an AFM on a Veeco Dimension 3100 system. Lift out of a sample for transmission electron microscopy (TEM) was achieved using a Thermo Fischer G3 Helios dual focused ion beam. To protect the QDs at the surface of the sample, a thick carbon layer was deposited using the electron beam followed by a Pt deposition with the electron beam. Final thinning was done at 2 kV with progressively lower currents (down to 48 pA) to avoid curtaining and amorphous damage. TEM was performed using a FEI Titan G2 8200. TEM was used to analyze the cross-sectional profile and coupled with energy-dispersive X-ray spectroscopy (EDS) to deduce the material composition of the QDs. The composition profile was also examined using secondary ion mass spectrometry (SIMS) measurements carried out by Evans Analytical Group (EAG).

**Optical characterization:** Temperature-dependent PL spectra of the densely packed GaAsSb QDs were measured from 4 K to 70 K in a Montana cryostat using a home-built confocal microscope equipped with a ×20 magnification objective lens (NA 0.4), integrated with a cryostat for temperature control. Excitation with 515 nm laser was provided with a pump power of 1 mW, and the emitted PL was analyzed using a Czerny–Turner spectrometer.

**First-principles calculations:** Electronic structure calculations of the As-for-Sb exchange reactions on the Sb-terminated GaSb (100) surface were carried out with the Quantum ESPRESSO package.[32] Reaction energies and nudged-elastic-band barriers are reported in Section S1 of the Supporting Information, together with the full computational parameters.

## Results and Discussion

The formation of GaAsSb QDs through the displacement etch technique is governed by the thermodynamically favorable reaction between $As_2$ molecular flux and GaSb surfaces. The underlying chemistry involves two primary reaction pathways that drive the nanostructure formation process:

$$\text{Anion Exchange Reaction: } 2GaSb(s) + As_2(g) \rightarrow 2GaAs(s) + Sb_2(g)$$

$$\text{Isoelectronic AsSb Formation: } GaSb(s) + As_2(g) \rightarrow GaAs(s) + AsSb(g)$$

The displacement reaction that forms the light-emitting QDs is depicted schematically in Figure 1a. Despite the name, the displacement etch converts the GaSb surface to GaAsSb rather than removing material from it. Both reactions exhibit negative enthalpy changes, indicating their thermodynamic favorability under the MBE growth conditions employed. We evaluated both pathways explicitly with density functional theory (DFT) on an Sb-terminated GaSb (100) surface; the calculations are described in Section S1 of the Supporting Information. The single-substitution pathway, which desorbs AsSb, releases 6.57 eV, and the double-substitution pathway, which desorbs $Sb_2$, releases 6.45 eV. The corresponding nudged-elastic-band barriers are 0.86 eV and 0.84 eV. Neither the reaction energies nor the barriers differ enough to bias the surface toward one pathway, so both should be thermally accessible under the same growth conditions and should proceed with comparable probability. An immediate consequence is that the morphological evolution we describe below cannot originate in the initial surface exchange step itself and must instead be set by the subsequent adatom kinetics. The anion exchange reaction and the isoelectronic compound formation represent a direct substitution mechanism where arsenic atoms displace antimony atoms from the GaSb lattice, forming GaAs while releasing volatile $Sb_2$ or AsSb species, respectively. The relative contribution of these two pathways depends critically on the $As_2$ exposure conditions, including molecular flux, exposure duration, and substrate temperature. The etch process creates a compositional gradient between the existing GaSb and the newly formed GaAs, resulting in GaAsSb QDs. At moderate exposure the GaAsSb nanostructures remain isolated. Under higher exposure they grow larger and coalesce, and the untransformed material left between neighboring QDs appears as the array of nanovoids discussed below.

The GaSb surface is prepared as described in the Experimental Section and then exposed to an $As_2$ flux. Exposure here means the combination of $As_2$ beam equivalent pressure (BEP), exposure time and substrate temperature, since a lower substrate temperature raises the $As_2$ sticking coefficient and therefore the dose delivered at fixed flux and time. Systematic variation of that combined exposure gives three distinct regimes (Table 1):

Table 1: Displacement etch regimes on (100) GaSb.

| **$As_2$ Exposure** | **Process Parameter** | **Results** |
|---|---|---|
| Low Exposure Regime (QD Formation) | $As_2$ BEP: $7\times10^{-7}$ to $1\times10^{-6}$ Torr<br>Exposure time: 30-120 seconds | Isolated QDs; areal density $10^2$–$10^5$ $\mu m^{-2}$ |
| Intermediate Exposure Regime (Transition) | $As_2$ BEP: $1\times10^{-6}$ to $2\times10^{-6}$ Torr<br>Exposure time: 120-600 seconds | QD coalescence and directional elongation along $[\bar{1}10]$ |
| High Exposure Regime (Nanovoid Formation) | $As_2$ BEP: $>2\times10^{-6}$ Torr<br>Exposure time: >600 seconds | Nanovoid formation; 15–20 nm surface relief |

The substrate temperature during $As_2$ exposure critically influences the reaction kinetics through its effect on the $As_2$ sticking coefficient and surface diffusion rates.[33] Lower temperatures (480-500°C) enhance $As_2$ incorporation due to increased sticking probability, effectively amplifying the exposure even at constant flux and time. Conversely, higher temperatures (520-540°C) promote desorption and surface diffusion, requiring higher $As_2$ fluxes to achieve equivalent exposure levels. Following $As_2$ exposure, a 300-second antimony soak at $2\times10^{-6}$ Torr BEP serves multiple critical functions. The antimony flux stabilizes the newly formed nanostructures by providing a protective antimony-rich surface layer and prevents further arsenic incorporation.

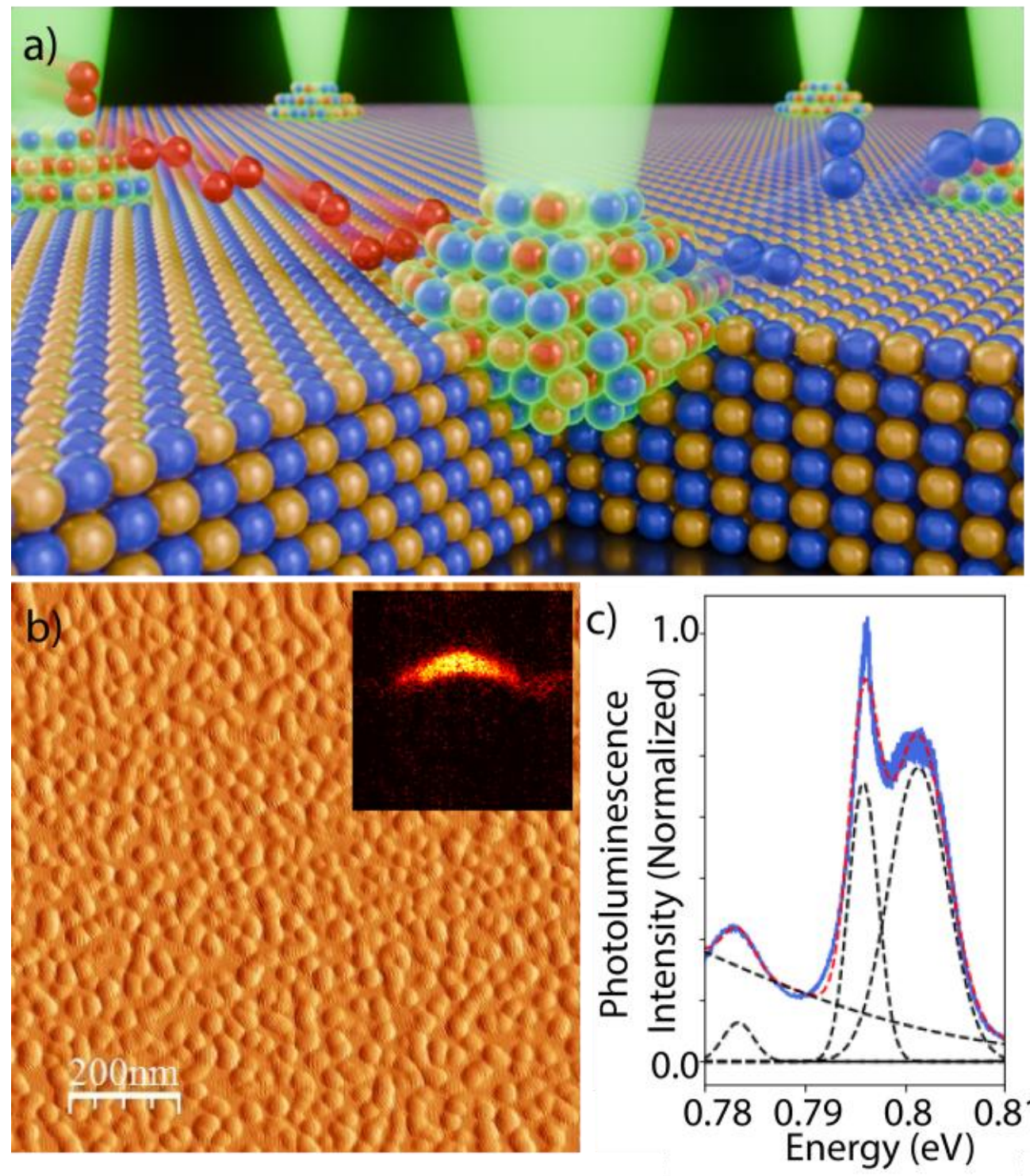

**Figure 1: GaAsSb QDs:** a) Schematic representation of the etch displacement reaction leading to the formation of GaAsSb (emitting light – see green colored regions) with red-colored As atoms displacing blue-colored Sb atoms. b) 1 × 1 µm atomic force microscopy (AFM) scan of As2-displaced nanostructures on the GaSb surface, 200 nm scale bar with an inset transmission electron microscopy (TEM) image of a single QD with black to yellow intensity correlating with %As incorporation. c) Photoluminescence (PL) intensity measured from an ensemble of GaAsSb QDs at 4 K plotted in blue as function of emission energies. The black dashed lines represent the Gaussian fits to the individual peaks which cumulatively (red-dashed lines) reproduce the measured spectrum.

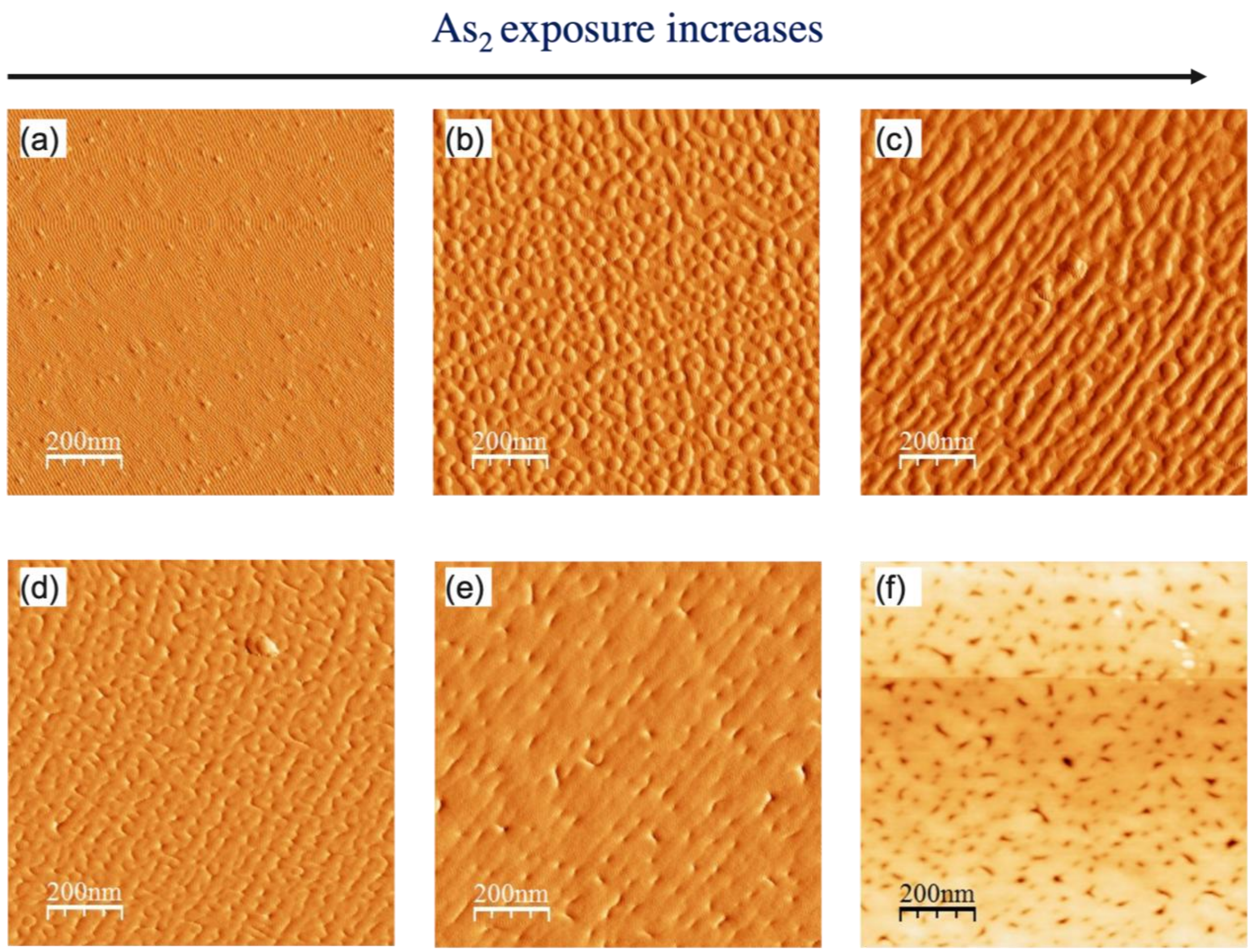

**Figure 2: Surface morphology changes with $As_2$ changes:** 1 × 1 µm AFM scans of (100) GaSb surfaces after increasing $As_2$ exposure (a–f). Exposure combines $As_2$ BEP, exposure time and substrate temperature (Table 1).

The $As_2$ exposure sets the QD areal density across three orders of magnitude. Figure 2 shows AFM scans of surfaces grown at different substrate temperatures with different $As_2$ BEPs and exposure times, spanning the low exposure regime and beyond it. Because $As_2$ BEP, exposure time and substrate temperature were co-varied between growths, no single parameter orders the series; the panels are arranged by increasing combined exposure, and Table 1 gives the parameter ranges spanned by each regime. At the lowest $As_2$ exposures (BEP ~$7\times10^{-7}$ Torr, 30-60 seconds), the displacement reaction occurs at isolated nucleation sites, creating sparse QD arrays with areal densities of order $10^2$ $\mu m^{-2}$ (Figure 2a). As the exposure increases, more nucleation sites become active and the areal density rises to ~$10^5$ $\mu m^{-2}$ (Figure 2b). Because the two surface exchange reactions are essentially equivalent (Section S1), the increase in density is not a change of reaction pathway; it follows from the larger number of arsenic atoms delivered to, and retained on, the surface, which raises the probability that a supercritical nucleus forms at any given site.

Individual QDs exhibit typical lateral dimensions of 20-30 nm and heights of 2-4 nm (Figure 1b), comparable in lateral extent to high-quality S-K InAs QDs.[34] The aspect ratio remains relatively constant across the density range, suggesting that the growth mechanism maintains consistent three-dimensional confinement characteristics regardless of areal density. Given that the displacement etch yields QDs with the structural characteristics of S-K QD, this means a three-dimensional growth mode can be triggered by a group-V exposure alone, without lattice-mismatch strain and without the etch-and-infill step that is required by local droplet etching. The rest of the paper establishes what these structures are and what they emit: we resolve their composition and crystal quality in cross-section, then measure their emission and its temperature dependence.

The $As_2$ etch of the (100) GaSb surface can be driven further to produce nanostructures beyond QDs. In the intermediate exposure regime (Table 1), QD coalescence and directional elongation occur preferentially along the $[1\bar{1}0]$ crystallographic direction (Figure 2c,d,e). This anisotropy reflects the direction-dependent surface diffusion of adatoms and the underlying crystal symmetry of the (100) GaSb surface, as established for the isostructural III–V (100) surface,[35,36] and it offers a route to aligned QD arrays for polarization-sensitive applications. Increasing the $As_2$ exposure into the high exposure regime (Table 1, Figure 2f) causes nearly complete coalescence of the nanostructures, leaving an array of nanovoids at the surface and approaching two-dimensional film growth under group-V flux alone. This morphology most closely resembles the surface reported for $As_2$ displacement of (100) GaSb by Huang et al.,[27] although we observe a crystallographic orientation absent from their results. We attribute the difference to operating in a regime where GaAsSb nanostructures coalesce, rather than one of strict material removal.

A single process step therefore gives access to isolated QDs, to quasi-one-dimensional structures, and to nanovoid arrays. The growth model can be interpreted by segmenting the process into (4) phases: (1) the onset of 3-dimensional growth driven by altered surface energy (As-for-Sb anion exchange) and compositional strain (formation of GaAsSb) (2) the density of the QDs increases with increasing arsenic exposure because this leads to an increase in the anion exchange rate and chemical supersaturation (3) At higher exposure, the anisotropic surface diffusion and reconstruction-dependent energy dynamics leads to elongation of the nanostructures along specific crystalline directions and (4) coalescence and/or redistribution of material due to volume contraction (GaSb-to-GaAs conversion) can deplete the material between the nanostructures and result in pits or voids.

For structural/compositional and optical analysis, the QDs (at $10^5/\mu m^{-2}$ areal density) are embedded in a 200 nm-thick $Al_{0.5}Ga_{0.5}Sb$ matrix. The QDs are formed by displacing a 4 nm GaSb quantum well (QW) grown in the middle of the AlGaSb matrix. For some samples, the etch process is repeated at the top surface (with a second 4 nm GaSb QW) for AFM measurements.

Cross-sectional TEM and EDS reveal the detailed structure and composition of the displacement-etched QDs. High-angle annular dark-field scanning TEM (HAADF-STEM) images confirm the formation of coherent three-dimensional QDs without observable defect propagation into the surrounding $Al_{0.5}Ga_{0.5}Sb$ barrier (Figure 3a). EDS analysis indicates that the QDs possess a compositionally graded structure with a GaAsSb core containing <50% antimony and a thin GaSb interfacial layer (Figure 3b). The line profile of Figure 3b is taken along the growth axis through the center of the QD marked in Figure 3a. Secondary ion mass spectrometry (SIMS) measurements (Figure S2, Supporting Information), despite being at the resolution limit for the QD layer, corroborate the EDS findings by showing a localized reduction in antimony concentration coincident with arsenic incorporation. The absence of extended defects at the QD interfaces, as confirmed by high-resolution TEM, indicates that the displacement reaction preserves the crystalline quality of the host material, comparable to that routinely obtained for S-K InAs QDs.[34] Interface defects act as non-radiative recombination centers,[37,38] so their absence is what keeps the internal quantum efficiency usable in a device. The composition map of Figure 3c resolves the same gradient in both the buried and the surface QDs, and shows three-dimensionally isolated regions of GaAsSb at sizes comparable to S-K QDs.

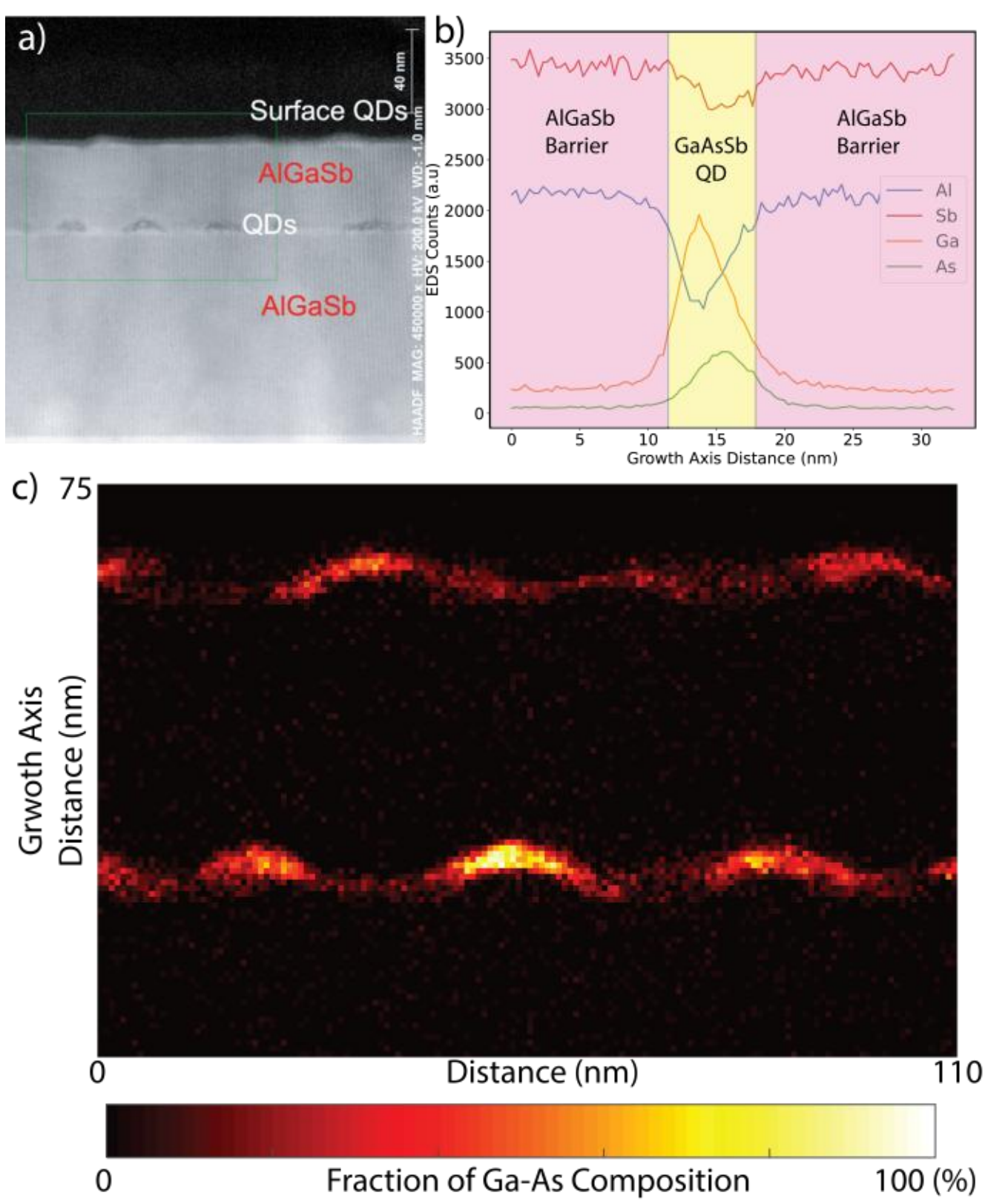

**Figure 3 Structural Analysis of the GaAsSb quantum dot (QD)**: a) cross-sectional HAADF-STEM image of the structure showing the spatially separated QDs sandwiched between the two AlGaSb layers along with the top surface QDs grown for AFM measurements. b) EDS profiles for the $As_2$ etched QD in terms of composition average over multiple slices near the quantum well (QW) centered at 15 nm along the growth axis. c) Compositional map derived from the same cross-sectional EDS dataset, showing the GaAsSb composition of the buried and surface QDs.

PL measurements were conducted to evaluate the optical properties of the displacement-etched GaAsSb QDs. The measurements were performed over a 4–70 K range using a 515 nm pump excitation and an InGaAs spectrometer (See method section for more details).

The QDs were embedded between $Al_{0.5}Ga_{0.5}Sb$ barrier layers, which provide the carrier confinement. The PL spectrum (Figure 4a) resolves three emission features: a band between 1540 and 1550 nm, a peak at 1558 nm, and a weaker feature near 1580 nm, all close to 0.80 eV and all within the telecommunications C-band. Based on the structural analysis showing that the QDs consist of a GaAsSb composition with <50% Sb content and an underlying thin GaSb layer, we observe that a reference sample mimicking the growth process except for the QD-formation process (4 nm GaSb QW sandwiched in AlGaSb without an $As_2$ etch) showed no sharp emission near 1550 nm; instead a single broad feature near 1570 nm, weaker by an order of magnitude, was measured (Figure 4b). This confirms that the emission in Figure 4a originates from the QDs. The observation of emission near 1550 nm for the GaAsSb QDs suggests a match with the C-band emission that corresponds to the low-loss transmission window of silica optical fibers,[39] making these emitters relevant for fiber-optic and telecommunication applications, provided the ensemble emission can be filtered or optimized down to a single narrow peak.

Due to the high areal density of the QDs, the detected signal originates from an ensemble rather than a single emitter, and the spread of QD sizes produces significant inhomogeneous broadening. As shown in Figure 4a, the PL spectra feature a prominent emission peak at 1558 nm (0.796 eV) at low temperatures, which decreases in intensity and broadens as temperature increases. To resolve the contribution from individual transitions within the broadened ensemble spectrum, we fitted the data with a sum of Gaussian functions, as illustrated in Figure 1c, where the measured spectrum (blue), the fit (red dashed line), and individual Gaussian components (black dashed lines) are overlaid. The extracted temperature dependence of the PL properties at the peak marked by the arrow in Figure 4a is summarized in Figure 4c,d. The peak PL intensity (Figure 4c) drops sharply with increasing temperature, indicating strong thermal quenching due to nonradiative recombination and carrier escape.[40] Meanwhile, the linewidth (Figure 4d) that remains nearly constant below ~30 K exhibits a marked increase above ~40 K, consistent with enhanced homogeneous broadening from phonon interactions and thermal carrier redistribution among QDs.[41,42] Both the intensity quenching and the linewidth broadening set in over the same temperature range (~30–40 K), consistent with thermally activated carrier escape and phonon-assisted redistribution as the common origin.[42] The corresponding pump-power series at 4 K (Figure S3, Supporting Information) shows the higher-energy shoulder growing with excitation, as expected for state filling of the ensemble, and establishes the excitation range over which the linewidth is power-independent.

**Implications for device use:** At areal densities on the order of 1 $\mu m^{-2}$, a diffraction-limited spot at 1.55 µm covers only a handful of QDs, which is the regime in which single-dot spectroscopy and single-photon emission become accessible When approaching areal densities on the order of $10^5$ $\mu m^{-2}$, the ensemble is dense enough to supply modal gain in a waveguide. Both ends of that areal density range are attainable by simply tuning the $As_2$ exposure on a given GaSb surface. Because the QDs sit in $Al_{0.5}Ga_{0.5}Sb$ on GaSb without extended defects, non-radiative recombination at the QD interface does not limit emission quantum efficiency, and the material is compatible with the epitaxial stack already used for mid-infrared sources and detectors on this platform. We identify two aspects of the material performance to improve before integration into a photonic device: narrowing or spectrally filtering the ensemble linewidth, and increasing the operating temperature above the current 4–70 K operating window. Single-dot spectroscopy and room-temperature measurements are the next steps toward photonic device integration.

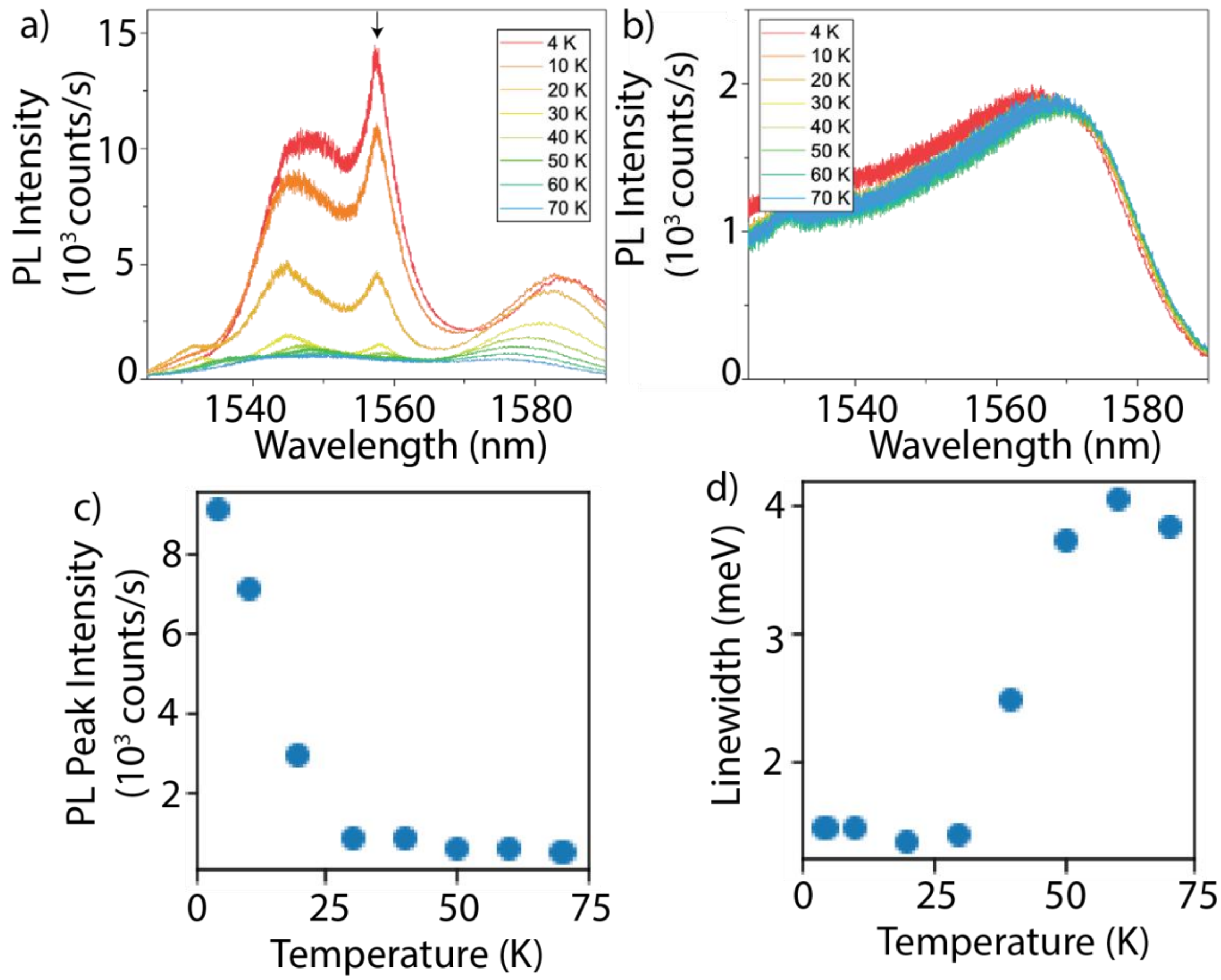

**Figure 4: Temperature dependent PL of the GaAsSb QDs**: a) Emission spectra at different temperatures at fixed optical pump power for the GaAsSb QD sample; the arrow marks the peak analyzed in c) and d). b) The same measurement on the reference sample, which contains all of the epitaxial layers except the QDs. Note the fivefold smaller intensity scale. c) and d) give the peak intensity and the linewidth of the arrow-marked peak as a function of temperature.

## Conclusions

We have demonstrated a displacement etch technique for epitaxial QD formation that addresses fundamental limitations of existing growth methods. A single exposure parameter sets the areal density across three orders of magnitude ($10^2$ to $10^5$ μm$^{-2}$), with material flexibility in principle comparable to droplet epitaxy and processing simplicity approaching S-K growth. The technique operates through thermodynamically favorable displacement reactions between $As_2$ molecular flux and GaSb surfaces, creating coherent three-dimensional QDs without extended defects, and spanning morphologies from isolated QDs through quasi-one-dimensional structures to arrays of nanovoids. First-principles calculations show that the two candidate surface exchange reactions are energetically and kinetically near-degenerate, so this morphological control is governed by adatom kinetics rather than by a switch in reaction pathway. Structural analysis confirms the formation of compositionally graded GaAsSb QDs with <50% antimony content, while temperature-dependent PL measurements demonstrate telecommunications wavelength emission at ~1550 nm with characteristic QD behavior. Displacement-etch epitaxy therefore delivers device-relevant QD densities and C-band emission on a GaSb platform from a single process step, and gives access to isolated QDs, aligned quasi-one-dimensional structures and nanovoid arrays with trivial changes to the growth recipe. Future work will focus on optimizing the emission properties through composition and confinement engineering, developing electrically-pumped devices, and exploring applications in quantum communication systems and integrated photonic circuits. The ability to precisely control QD density from sparse arrays suitable for single-photon sources to high-density ensembles for laser gain media, combined with emission in the telecommunications C-band, positions these structures as promising candidates for next-generation quantum photonic and integrated optical devices.

## Acknowledgements

This work was performed, in part, at the Center for Integrated Nanotechnologies, an Office of Science User Facility operated for the U.S. Department of Energy (DOE) Office of Science. Sandia National Laboratories is a multimission laboratory managed and operated by National Technology & Engineering Solutions of Sandia, LLC, a wholly owned subsidiary of Honeywell International,

Inc., for the U.S. DOE's National Nuclear Security Administration under contract DE-NA-0003525. The views expressed in the article do not necessarily represent the views of the U.S. DOE or the United States Government. S.J.A. acknowledges funding from Sandia's LDRD Program

Supplementary Information

# Epitaxial Displacement etched GaAsSb Quantum dots emitting at Telecom Wavelengths

Sadhvikas Addamane[1*], Sanghyeok Park[1], Quinn T. Campbell,[2] Evan M. Anderson,[2] Thomas Rotter[3], Darryl Shima[3], Julia Deitz[2], Paul Kotula[2], Ganesh Balakrishnan[3], Oleg Mitrofanov[4], Igal Brener[1] and Prasad P. Iyer[1*]

*[1]Center for Integrated Nanotechnologies, Sandia National Laboratories, Albuquerque, NM 8712, USA*
*[2]Sandia National Laboratories, Albuquerque, NM 8712, USA. [3]Center for High-Technology Materials, University of New Mexico, Albuquerque, NM, 87106. [4]University College London, Electronic and Electrical Engineering, London WC1E 7JE, U.K.*
*saddama@sandia.gov, *ppadma@sandia.gov

## S1. First-Principles Calculations of Surface Chemical Reactions

We used electronic structure calculations to determine the thermodynamic and kinetic energies of the two surface reactions considered in the main text:

$$2GaSb(s) + As_2(g) \rightarrow 2GaAs(s) + Sb_2(g)$$

$$GaSb(s) + As_2(g) \rightarrow GaAs(s) + AsSb(g)$$

Under growth conditions both reactions take place at the surface of the GaSb substrate. We therefore model the Sb-terminated GaSb (100) surface using density functional theory (DFT), with the atomic structure shown in Figure S1a. We start from an $As_2$ molecule placed above the surface and model two possible endpoints: one in which both As atoms of the molecule are swapped for surface Sb atoms, forming an $Sb_2$ molecule (Figure S1b), and one in which only a single As atom swaps with a surface Sb, forming an AsSb molecule (Figure S1c).

Both endpoints are thermodynamically favorable and release nearly the same energy. The single-As swap is marginally more favorable, at 6.57 eV, compared with 6.45 eV for the double substitution. That difference is too small to bias the surface strongly toward either product. The kinetic barriers are likewise close: 0.84 eV for the two-atom swap and 0.86 eV for the single-atom swap. Because both the reaction energies and the barriers are so similar, the two reactions should occur with roughly equal probability and should be accessible at the same thermal activation. Both therefore contribute to the formation of GaAsSb QDs. This has a direct bearing on the main text: the change in morphology with increasing $As_2$ exposure is unlikely to originate in the initial surface exchange reaction, and is instead governed by the subsequent adatom kinetics and coalescence.

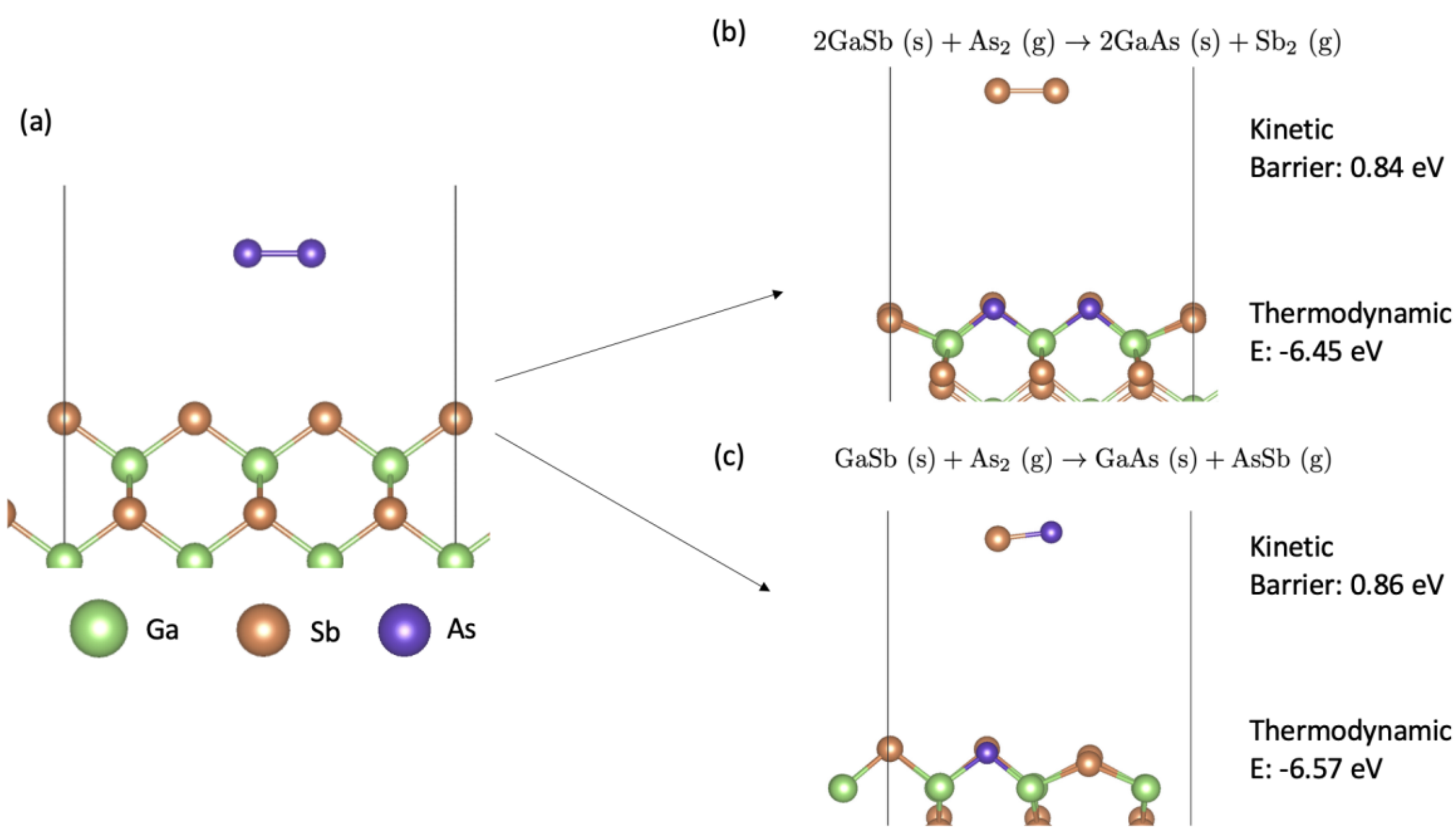

Figure S1: Atomic structure of a) the Sb-terminated GaSb (100) surface as an $As_2$ molecule approaches, b) the surface after both As atoms have swapped with surface Sb atoms, creating GaAs units at the surface and releasing an $Sb_2$ molecule, and c) the system in which only one As atom swaps at the surface, releasing an AsSb molecule. Ga atoms are shown in green, Sb in orange and As in purple.

**Computational methods:** Electronic structure calculations were performed in a slab geometry with the Quantum ESPRESSO package.[S1] We used the Perdew–Burke–Ernzerhof (PBE) exchange–correlation functional[S2] with norm-conserving pseudopotentials from the PseudoDojo repository.[S3] Kinetic energy cutoffs of 680 eV and 5440 eV were used for the plane-wave basis sets describing the Kohn–Sham orbitals and the charge density, respectively. The Brillouin zone was sampled with a 2×2×1 Monkhorst–Pack grid.[S4] Equilibrium structures were obtained by relaxing forces below 0.05 eV/Å. Reaction barriers were computed with the nudged elastic band (NEB) method with climbing image, as implemented in Quantum ESPRESSO.

**S2. Secondary Ion Mass Spectrometry (SIMS) of the QD and reference samples**

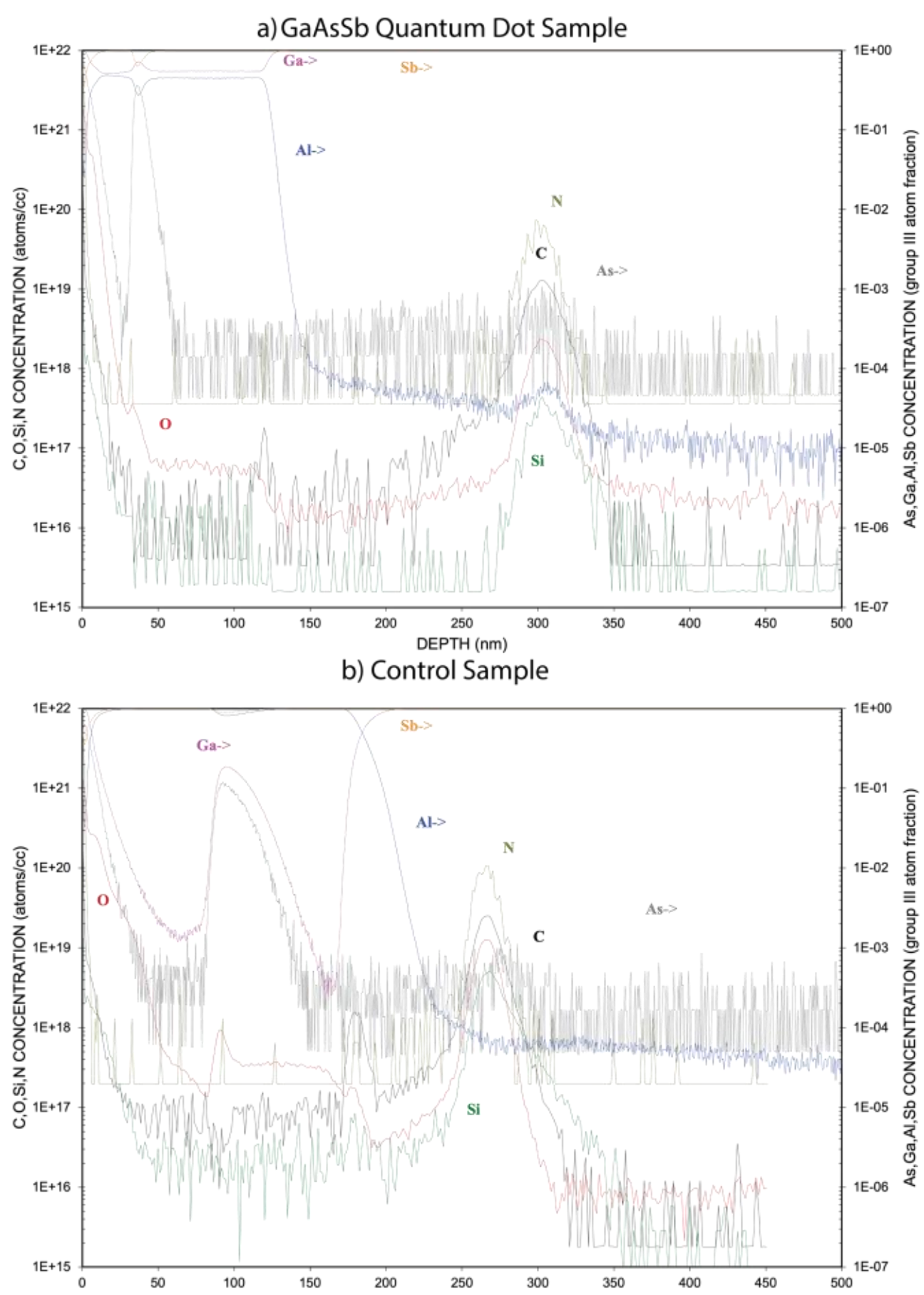

**Figure S2: SIMS depth profiles of the QD and reference samples.** Secondary-ion mass spectrometry elemental depth profiles (As, Sb, Al, Ga/Ge, O, N) for the QD sample (top) and the control (bottom). In the QD sample, a localized arsenic peak coincides with a dip in antimony at the dot-layer depth, consistent with As-for-Sb displacement; the control shows no comparable localized As incorporation. Profiles are at the depth resolution limit of the technique for this ~4 nm layer.

**S3. Pump-power dependence of QD emission**

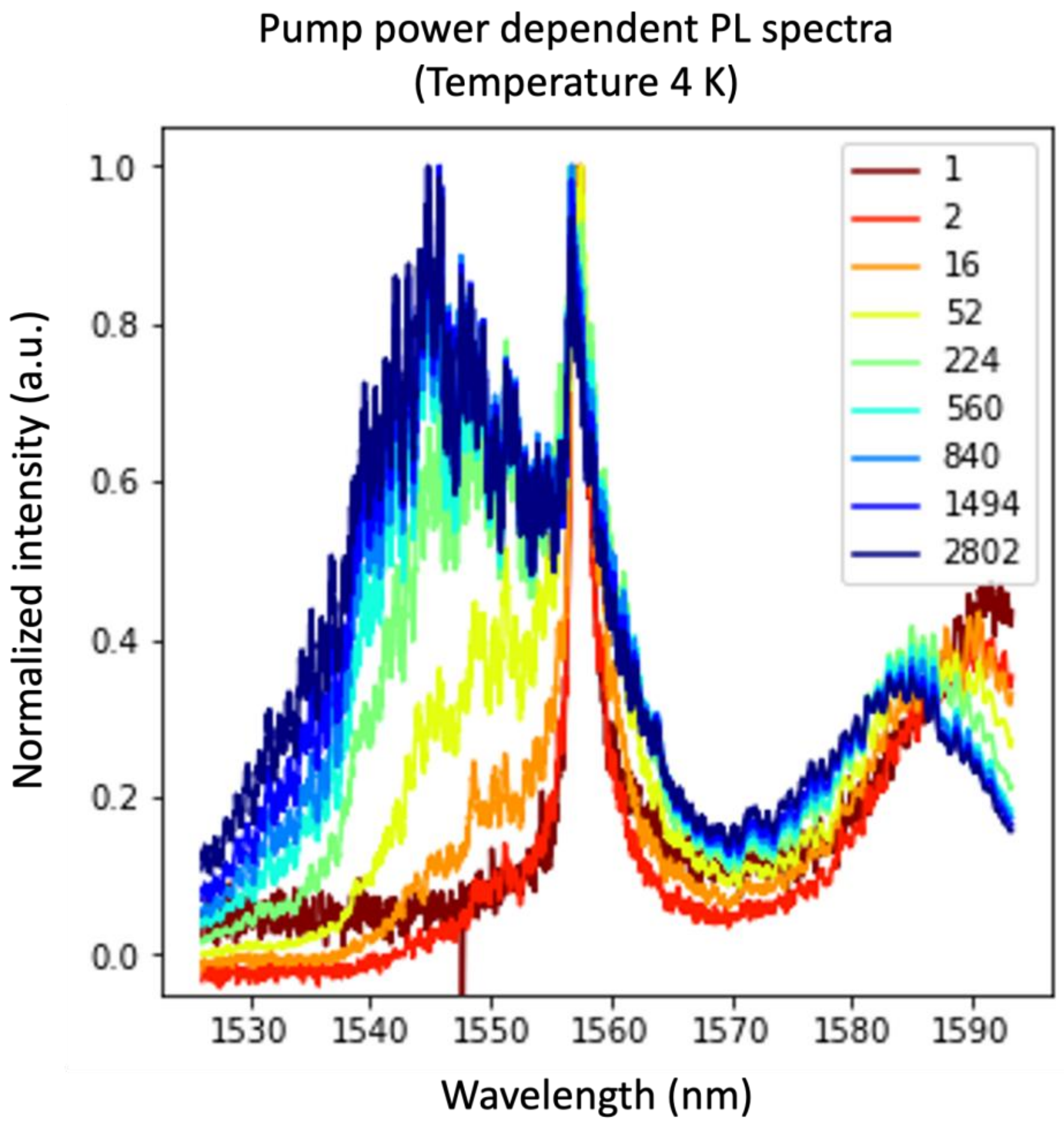

**Figure S3: Pump-power-dependent PL at 4 K.** Normalized PL spectra of the GaAsSb QD ensemble at 4 K as excitation power is increased (legend: 1–2802 W/cm$^2$). The 1558 nm ground-state peak dominates at low power; with increasing excitation a higher-energy shoulder (~1540–1550 nm) emerges and grows, indicating state filling of the QD ensemble. This series also establishes the excitation range over which the emission linewidth is power-independent.